\documentclass[fleqn,usenatbib]{mnras}

\usepackage[T1]{fontenc}
\usepackage{ae,aecompl}

\usepackage{newtxtext,newtxmath}

\usepackage{graphicx}
\usepackage{amsmath}
\usepackage{amssymb}
\usepackage{aas_macros}
\usepackage{color}
\usepackage{hyperref}
\defcitealias{Maltby_etal:2016}{M16}

\title[High-velocity outflows in PSBs at $z > 1$]{High-velocity outflows in massive post-starburst galaxies
at $z > 1$}
\author[D.~T.~Maltby et al.]
{David~T.~Maltby,$^{1}$\thanks{E-mail: david.maltby@nottingham.ac.uk}
 Omar~Almaini,$^{1}$ Ross~J.~McLure,$^{2}$ Vivienne~Wild,$^{3}$ James~Dunlop,$^{2}$
 \newauthor Kate~Rowlands,$^{4}$ William~G.~Hartley,$^{5}$ Nina~A.~Hatch,$^{1}$ Miguel~Socolovsky,$^{1}$
 \newauthor Aaron Wilkinson,$^{6}$ Ricardo~Amorin,$^{7,8}$ Emma~J.~Bradshaw,$^{1}$ Adam~C.~Carnall,$^{2}$
 \newauthor Marco~Castellano,$^{9}$ Andrea~Cimatti,$^{10,11}$ Giovanni~Cresci,$^{11}$ Fergus~Cullen,$^{2}$ 
 \newauthor Stephane~De~Barros,$^{12}$ Fabio~Fontanot,$^{13}$ Bianca~Garilli,$^{14}$ Anton~M.~Koekemoer,$^{15}$
 \newauthor Derek~J.~McLeod,${^2}$ Laura~Pentericci$^{9}$ and Margherita~Talia$^{10,16}$\\
$^{1}$School of Physics and Astronomy, University of Nottingham, University Park, Nottingham NG7 2RD, UK\\
$^{2}$Institute for Astronomy, University of Edinburgh, Royal Observatory, Blackford Hill, Edinburgh EH9 3HJ, UK\\
$^{3}$School of Physics and Astronomy, University of St Andrews, North Haugh, St Andrews KY16 9SS, UK\\
$^{4}$Department of Physics and Astronomy, Johns Hopkins University, Bloomberg Center, 3400 N.~Charles St., Baltimore, MD 21218, USA\\
$^{5}$Department of Physics and Astronomy, University College London, 3rd Floor, 132 Hampstead Road, London NW1 2PS, UK\\
$^{6}$Sterrenkundig Observatorium, Universiteit Gent, Krijgslaan 281 S9, 9000 Gent, Belgium\\
$^{7}$Departamento de F{\'i}sica y Astronom{\'i}a, Universidad de La Serena, Av.~Juan Cisternas 1200 Norte, La Serena, Chile\\
$^{8}$Instituto de Investigaci{\'o}n Multidisciplinar en Ciencia y Technolog{\'i}a, Universidad de La Serena, Ra{\'u}l Bitr{\'a}n 1305, La Serena, Chile\\
$^{9}$INAF--Osservatorio Astronomico di Roma, Via Frascati 33, 00078 Monte Porzio Catone, Italy\\
$^{10}$Dipartimento di Fisica e Astronomia, Universit{\`a} di Bologna, Via Piero Gobetti 93/2, 40129 Bologna, Italy\\
$^{11}$INAF--Osservatorio Astrofisico di Arcetri, Largo Enrico Fermi 5, 50125 Firenze, Italy\\
$^{12}$Observatoire de Gen{\`e}ve, Universit{\'e} de Genev{\`e}, Chemin des Maillettes 51, 1290 Versoix, Switzerland\\
$^{13}$INAF--Osservatorio Astronomico di Trieste, Via Giambattista Tiepolo 11, 34143 Trieste, Italy\\
$^{14}$INAF--Istituto di Astrofisica Spaziale e Fisica Cosmica di Milano, Via Bassini 15, 20133 Milano, Italy\\
$^{15}$Space Telescope Science Institute, 3700 San Martin Drive, Baltimore, MD 21218, USA\\
$^{16}$INAF--Osservatorio di Astrofisica e Scienza dello Spazio di Bologna, Via Piero Gobetti 93/3, 40129 Bologna, Italy
\vspace{-0.50cm}}

\begin{document}

\date{Accepted 2019 August 5. Received 2019 July 26; in original form 2019 February 1.
\vspace{-0.20cm}}

\pagerange{\pageref{firstpage}--\pageref{lastpage}} \pubyear{0000}

\maketitle

\label{firstpage}


\begin{abstract}
We investigate the prevalence of galactic-scale outflows in post-starburst (PSB) galaxies at high redshift
($1 < z < 1.4$), using the deep optical spectra available in the UKIDSS Ultra Deep Survey (UDS).  We use a
sample of $\sim40$ spectroscopically confirmed PSBs, recently identified in the UDS field, and perform a
stacking analysis in order to analyse the structure of strong interstellar absorption features such as
Mg\,{\sc ii} ($\lambda2800$\,\AA).  We find that for massive ($M_* > 10^{10}\rm\,M_{\odot}$) PSBs at
$z > 1$, there is clear evidence for a strong blue-shifted component to the Mg\,{\sc ii} absorption feature,
indicative of high-velocity outflows ($v_{\rm out}\sim1150\pm160\rm\,km\,s^{-1}$) in the interstellar medium.
We conclude that such outflows are typical in massive PSBs at this epoch, and potentially represent the
residual signature of a feedback process that quenched these galaxies.  Using full spectral fitting, we also
obtain a typical stellar velocity dispersion~$\sigma_*$ for these PSBs of $\sim200\rm\,km\,s^{-1}$, which
confirms they are intrinsically massive in nature (dynamical mass $M_{\rm d}\sim10^{11}\rm\,M_{\odot}$).
Given that these high-$z$ PSBs are also exceptionally compact ($r_{\rm e}\sim1$--$2\rm\,kpc$) and spheroidal
(S{\'e}rsic index~$n\sim3$), we propose that the outflowing winds may have been launched during a recent
compaction event (e.g.\ major merger or disc collapse) that triggered either a centralised starburst or
active galactic nuclei (AGN) activity.  Finally, we find no evidence for AGN signatures in the optical
spectra of these PSBs, suggesting they were either quenched by stellar feedback from the starburst itself, or
that if AGN feedback is responsible, the AGN episode that triggered quenching does not linger into the
post-starburst~phase.
\end{abstract}

\begin{keywords}
galaxies: high-redshift --- galaxies: ISM --- galaxies: kinematics and dynamics
\vspace{-0.50cm}
\end{keywords}

\section[]{Introduction}

\label{Introduction}

In the local Universe, there exists a clear bi-modality in the galaxy population with respect to optical
colour, star-formation characteristics and morphology \citep[e.g.][]{Strateva_etal:2001,
Schawinski_etal:2014}.  In general, massive galaxies tend to be red, passive and of early-type morphology,
while lower mass galaxies tend to be blue, star forming and of late-type morphology.  These two populations
form the {\em red-sequence} and {\em blue cloud}, respectively.  Significant evolution in this bi-modality
has been observed since $z > 2$, showing a rapid build-up of mass upon the red-sequence \citep[e.g.][]
{Bell_etal:2004, Cirasuolo_etal:2007, Faber_etal:2007, Brammer_etal:2011, Ilbert_etal:2013,
Muzzin_etal:2013}.  However, the principal drivers behind the required quenching of blue cloud galaxies at
high redshift remain uncertain and a topic of significant debate.

To account for the quenching of star formation at high redshift, several mechanisms have been proposed.  For
example, gas stripping processes \citep[e.g.][]{Gunn&Gott:1972}, morphological quenching
\citep{Martig_etal:2009}, shock heating of infalling cold gas by the hot halo \citep{Dekel&Birnboim:2006},
and an exhaustion of the gas supply \citep[e.g.][]{Larson_etal:1980}.  Other promising contenders include
feedback processes, where the outflowing superwinds generated by either an AGN or starburst can expel the
cold gas required for continuous star-formation \citep[e.g.][]{Silk&Rees:1998, Hopkins_etal:2005,
Diamond-Stanic_etal:2012}.  To prevent further gas accretion, and therefore keep star formation suppressed,
radio-mode AGN feedback may also be required \citep{Best_etal:2005, Best_etal:2006}.  In general, these
quenching mechanisms fall into two main categories: i)~those that lead to a rapid truncation of star
formation ({\em rapid quenching}), and ii)~those that prevent the accretion of new gas resulting in a more
gradual decline ({\em slow quenching}).  With respect to the dominant quenching route, recent studies have
indicated that slow quenching dominates in the local Universe \cite[e.g.][]{Peng_etal:2015}, while rapid
quenching (e.g.\ feedback processes) becomes increasingly more important at $z > 1$ \citep[e.g.][]
{Barro_etal:2013, Wild_etal:2016, Carnall_etal:2018, Belli_etal:2018}.

With respect to feedback processes, there is strong evidence for the galactic-scale outflows required to
quench high-$z$ galaxies.  At low- and high-redshift, outflows spanning a wide range of velocities have been
detected in both i)~AGN of various types \citep[e.g.][]{Hainline_etal:2011, Harrison_etal:2012,
Cimatti_etal:2013, Cicone_etal:2014, Talia_etal:2017}; and ii)~star-forming galaxies with no signs of AGN
activity, either in their UV/optical spectrum \citep[e.g.][]{Talia_etal:2012, Bradshaw_etal:2013,
Bordoloi_etal:2014, Talia_etal:2017} or X-ray properties \citep{Cimatti_etal:2013}.  These studies have also
revealed that such gaseous outflows are a multiphase phenomenon and exist in each of the
high-/low-ionisation, neutral and molecular gas phase of the interstellar medium \cite[ISM; e.g.][]
{Hainline_etal:2011, Cicone_etal:2014, Fluetsch_etal:2019, RobertsBorsani&Saintonge:2019}.

In order for feedback processes to quench a galaxy, strong outflows capable of expelling the gas reservoir
are required.  In the local Universe, strong outflows are only detected in starburst galaxies
\cite[e.g.][]{Heckman_etal:2000, Martin:2005, Heckman_etal:2015}, but at higher redshifts ($z>0.5$) such
outflows are more ubiquitous among the general star-forming population \citep[e.g.][]{Weiner_etal:2009,
Bradshaw_etal:2013, Rubin_etal:2014, Du_etal:2018}.  Particularly, strong outflows
($v_{\rm out}\sim1000\rm\,km\,s^{-1}$) have also been detected in both star-forming AGN at $z > 1$
\citep[e.g.][]{Hainline_etal:2011, Harrison_etal:2012, Talia_etal:2017}, as well as at $z\sim0.6$ in massive
compact star-forming/starburst galaxies with no signs of AGN activity \citep[][]{Geach_etal:2014,
Sell_etal:2014}.  Taken together, these results indicate that both AGN and starburst-driven winds are capable
of driving the strong outflows required for rapid quenching.  For starburst galaxies at $z < 1$, recent
studies have also reported that the outflow strength (i.e.~velocity) depends on stellar mass, star-formation
rate (SFR) and, in particular, SFR density \citep[e.g.][]{Heckman_etal:2015, Heckman_etal:2016}, suggesting
feedback from star-formation can be a principal driver of strong galactic-scale outflows.  However, although
AGN and starburst-driven outflows clearly represent a promising mechanism to explain the quenching of
star-formation at high redshift, observationally a direct causal link to quenching remains elusive.

To establish the role of outflows as a quenching process, it is\break useful to consider galaxies that have been
recently quenched.  The\break rare population of post-starburst (PSB) galaxies provide an ideal example, as they
represent systems that have experienced a major burst of star formation that was rapidly quenched at some
point during the last Gyr.  These galaxies are identified spectroscopically from the characteristic strong
Balmer absorption lines related to an enhanced A-star population, combined with a general lack of strong
emission\break\mbox{lines\,\citep{Dressler&Gunn:1983, Wild_etal:2009}.  At\,intermediate\,red-}\break\mbox{shifts ($z\sim0.6$),
particularly strong outflows} ($v_{\rm out} > 1000\rm\,km\,s^{-1}$)\break have been detected in the most luminous
PSBs \citep[$M_{\rm B}\sim-23.5$;][]{Tremonti_etal:2007}, potentially representing the residual outflow\break from
a quenching event.  However, such galaxies are extremely rare,\break and may not represent the typical evolutionary
path of red-sequence\break galaxies at this epoch.  More modest outflows ($v_{\rm out}\sim200\rm\,km\,s^{-1}$)\break have
also been observed in less luminous PSBs at $0.2 < z < 0.8$ \citep[$M_{\rm B}\sim-21$;][]{Coil_etal:2011},
but it is unclear whether such outflows \mbox{are sufficient to actually quench star-formation.  At higher
redshifts} ($z > 1$), where we observe a rapid build-up of mass upon the red-sequence, recent evidence
suggests that rapid quenching becomes increasingly more important \citep[e.g.][]{Barro_etal:2013,
Carnall_etal:2018, Belli_etal:2018} and that a large fraction of massive galaxies
($M* > 10^{10.5}\rm\,M_{\odot}$) will experience a PSB phase \citep{Wild_etal:2016, Belli_etal:2018}.
However, the nature of outflows in massive PSBs \mbox{at this epoch has, until now, been largely unexplored.}

Until recently, very few PSBs had been spectroscopically identified at high redshift ($z > 1$).  However,
significant progress was made by \cite{Maltby_etal:2016}, when photometric PSB candidates identified
using the \cite{Wild_etal:2014} {\em `supercolour'} technique were targeted for follow-up spectroscopy.  This
led to $>20$ high-$z$ PSBs being identified within the field of the Ultra Deep Survey (UDS; Almaini et al.,
in preparation).  In this paper, we use the deep optical spectra of \cite{Maltby_etal:2016}, plus additional
spectra obtained more recently within the UDS field (see Section~\ref{Data}), to determine the prevalence of
outflows in these galaxies.  To achieve this, we perform a stacking analysis and analyse the structure of
strong interstellar absorption features such as Mg\,{\sc ii} ($\lambda2800$\,\AA).  This is achievable for
the first time with our large sample of high-$z$ PSB spectra.

The structure of this paper is as follows.  In Section~\ref{Data}, we provide a brief description of the UDS
data and spectroscopy upon which this work is based, including details of our PSB spectra and stacking
procedure.  In Section~\ref{velocity dispersions}, we perform full spectral fits on our stacked spectra in
order to measure their typical stellar velocity dispersions~$\sigma_*$, while in Section~\ref{Outflows} we
describe the method used for detecting outflows from the Mg\,{\sc ii} absorption feature and present our
findings for high-$z$ PSBs.  Finally, we draw our conclusions in Section~\ref{Conclusions}.  Throughout this
paper, we use AB magnitudes and adopt a cosmology of $H_{0} = 70\rm\,km\,s^{-1}\,Mpc^{-1}$,
$\Omega_{\Lambda} = 0.7$ and $\Omega_{\rm{m}} = 0.3$.

\section[]{Description of the Data}

\label{Data}

\subsection[]{The UDS: photometric and spectroscopic data}

\label{The UDS}

This study makes use of the deep photometric data from the UDS (Almaini et al., in
preparation).\footnote{http://www.nottingham.ac.uk/astronomy/UDS/}  This survey represents the deepest
component of the UKIRT (United Kingdom Infra-Red Telescope) Infrared Deep Sky Survey \cite[UKIDSS;][]
{Lawrence_etal:2007} and comprises extremely deep UKIRT {\em JHK} photometry, covering an area of
$0.77\rm\deg^{2}$.  For this study, we make use of the eighth UDS data release (DR8) where the limiting
depths are $J = 24.9$; $H = 24.4$ and $K = 24.6$ (AB; $5\sigma$ in $2\rm\,arcsec$ apertures).  The UDS is
also complemented by extensive multiwavelength observations.  These include deep-optical {\em BVRi$'$z$'$}
photometry from the Subaru--{\em XMM-Newton} Deep Survey \citep[SXDS;][]{Furusawa_etal:2008}, mid-infrared
observations ($3.6$ and $4.5\rm\,\mu{m}$) from the {\em Spitzer} UDS Legacy Program (SpUDS; PI: Dunlop) and
deep $u'$-band photometry from MegaCam on the Canada--France--Hawaii Telescope (CFHT).  The extent of the UDS
field with full multiwavelength coverage (optical--mid-infrared) is $\sim0.62\rm\,deg^{2}$.  For a complete
description of these data, see \cite{Hartley_etal:2013} and \cite{Simpson_etal:2012}.  In this work, where
appropriate, we use the photometric redshifts and SED-derived stellar masses described in
\cite{Simpson_etal:2013}.  We also use the galaxy $K$-band structural parameters (effective
radius~$r_{\rm e}$; S{\'e}rsic index~$n$) described in \cite{Almaini_etal:2017}.

Extensive deep optical spectroscopy is also available within the UDS field.  These data are provided by
several spectroscopic programmes.  The largest sample was obtained by UDSz, the spectroscopic component of
the UDS (ESO large programme 180.A-0776), which used both the VIMOS and FORS2 instruments on the ESO VLT to
obtain low-/medium-resolution optical spectra for $>3500$ galaxies \citep[$R_{\rm\,VIMOS}\sim200$ and
$R_{\rm\,FORS2}\sim660$; exposures of $2.6$--$4.5$\,h and $5.5$\,h, respectively; see][]{Bradshaw_etal:2013,
Mclure_etal:2013}.  This is complemented by the spectroscopic follow-up of the \cite{Wild_etal:2014,
Wild_etal:2016} sample of photometrically selected PSBs (ESO programme 094.A-0410; hereafter M$_{16}$), which
provides $\sim100$ medium-resolution optical spectra from VIMOS \citep[$R\sim580$; exposures of $4$\,h;
see][]{Maltby_etal:2016}.  Finally, the VANDELS spectroscopic survey (ESO programme 194.A-2003) also targets
the UDS field, providing an additional $\sim780$ medium-resolution VIMOS spectra ($R\sim580$; exposures of
$20$\,h or $40$\,h), mainly at $z > 2$, from the second data release \citep[DR2;][]{McLure_etal:2018,
Pentericci_etal:2018}.  From these datasets, $>2300$ secure spectroscopic redshifts~$z_{\rm spec}$ are
available, all of which were determined via {\sc ez} \citep{Garilli_etal:2010}, which uses a
cross-correlation of spectral templates.  Optimal solutions were also confirmed using spectral line
identification in {\sc sgnaps} \citep{Paioro&Franzetti:2012}.  For further details on these spectroscopic
redshifts, see the relevant data papers.  In this work, we use these $z_{\rm spec}$ to shift the individual
galaxy spectra to their respective rest-frame (i.e.\ systemic frame).

In this spectroscopic study, we focus specifically on the redshift interval $1 < z < 1.4$ (see
Section~\ref{Sample selection}).  Within the UDS (DR8) there are $\sim9000$ $K$-band selected galaxies within
this redshift range above the $95$~per~cent mass-completeness limit of the survey
(\mbox{$M_*\sim10^{9.5}\rm\,M_{\odot}$} at $z\sim1$; as determined using the method of
\citealt{Pozzetti_etal:2010}).  Of these galaxies, $\sim6$~per~cent have available optical spectra provided
by the datasets above.  These spectra evenly sample both the $M_*$ and redshift distribution of the parent
photometric sample.

\subsection[]{Post-starburst galaxies in the UDS}

\label{Sample selection}

In general, PSBs are spectroscopically identified from the presence of strong Balmer absorption lines
(e.g.\ H\,$\delta$ $\lambda4102$\,\AA), combined with a general lack of strong emission lines
\citep{Dressler&Gunn:1983, Wild_etal:2009}.  Therefore, to identify PSBs in the UDS, we use all the available
optical spectroscopy (see Section~\ref{The UDS}) and apply the following criteria, where applicable: i)~an
equivalent width in H\,$\delta > 5$\,\AA\, \citep[a general PSB diagnostic; e.g.][]{Goto:2007}; and ii)~an
equivalent width in [O\,{\sc ii}] $> -5$\,\AA\ (a standard threshold to remove galaxies with significant
on-going star formation; see e.g.\ \citealt{Tran_etal:2003, Poggianti_etal:2009, Maltby_etal:2016}).  We note
that, while a cut on [O\,{\sc ii}] emission is necessary to avoid the contamination of\break our samples by
star-forming galaxies, this criterion will also remove\break some genuine PSBs that host significant AGN activity
\citep[][]{Yan_etal:2006}.  \mbox{The\,equivalent\,width\,($W_{\lambda}$)\,of\,a\,spectral\,line\,is\,defined\,as}
\begin{equation}
{W_{\lambda}} = \int_{\lambda_{1}}^{\lambda_{2}}\!1-F(\lambda)/F_{\rm c}(\lambda)\,{\rm{d}}\lambda,
\end{equation}
where $F(\lambda)$ is the spectral flux and $F_{\rm c}(\lambda)$ is the continuum flux.  To determine the
rest-frame equivalent width ($W_{\lambda}$), we use a non-parametric approach based on that used by previous
works \citep[e.g.][]{Goto_etal:2003, Maltby_etal:2016}.  First, $z_{\rm{spec}}$ is used to transform the
spectrum into the galaxy's rest-frame.  Then continuum flux is estimated across the relevant feature
(i.e.\ H\,$\delta$, [O\,{\sc ii}]) using a linear interpolation between the continuum measured in narrow
intervals on either side.  These intervals are chosen to ensure a lack of significant absorption/emission
lines and the continuum is modelled by a linear regression that includes both intervals, weighted by the
inverse square error in the flux.  A $3\sigma$ rejection to deviant points above/below an initial continuum
model is also used to minimise the effect of noise.  Finally, $W_{\lambda}$ is determined using the ratio
$F(\lambda)/F_{\rm c}(\lambda)$ across an interval ($\lambda_1$--$\lambda_2$) that encapsulates the feature
of interest \citep[see][for more details]{Maltby_etal:2016}.  For each spectrum, the uncertainty in
$W_{\lambda}$ ($1\sigma$) is determined from the $W_{\lambda}$ variance between $1000$ simulated spectra
generated by using the $1\sigma$ flux errors to add suitable Gaussian noise.  Typical uncertainties in both
$W_{\,\rm H\delta}$ and $W_{\,\rm [O\,II]}$ are $\pm1$\AA\ ($\sim 15$ per cent).

\begin{figure}
\includegraphics[width=0.43\textwidth]{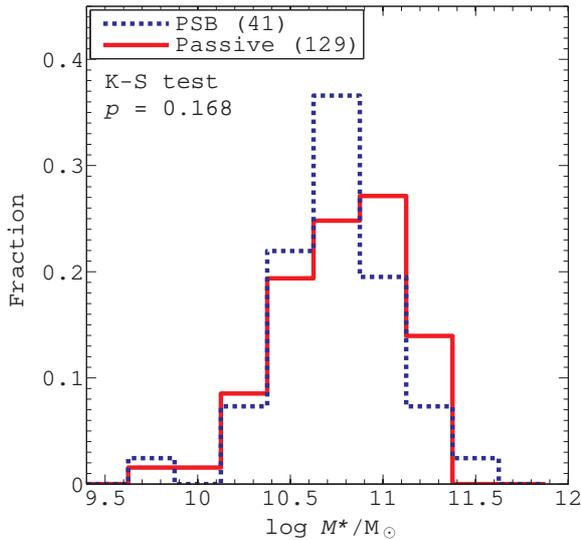}
\centering
\vspace{-0.1cm}
\caption{\label{stellar mass}The distribution of stellar mass $M_*$ for our high redshift
\mbox{($1 < z < 1.4$)} PSB and passive galaxy spectra.  In both cases, these galaxies are typically of high
stellar mass ($M_* > 10^{10}\rm\,M_{\odot}$) and a Kolmogorov--Smirnov (K--S) test reveals no significant
difference between their $M_*$ distributions ($p = 0.168$).  Relevant sample sizes are shown in the legend.}
\end{figure}

In this study, PSB classification depends on the assessment of both H\,$\delta$ ($\lambda4102$\,\AA) and
[O\,{\sc ii}] ($\lambda3727.5$\,\AA).  Using the available data (see Section~\ref{The UDS}), this requirement
restricts our PSB classification to $z\lesssim1.4$, where these features are within the reach of our
spectroscopy.  This limits our analysis to $502$ spectra at $z > 1$, for which a spectroscopic PSB assessment
is possible ($386$~UDSz, $32$~M$_{16}$ and $84$~VANDELS spectra).  Applying our PSB criteria to these spectra
(i.e.\ $W_{\,\rm H\delta} > 5$\,\AA, $W_{\,\rm [O\,II]} > -5$\,\AA), we obtain a final sample of $41$
spectroscopically classified PSBs at $1 < z < 1.4$ (see Table~\ref{sample properties}).  For comparison to
these PSBs, we also identify older passive systems (i.e.~$W_{\,\rm H\delta} < 5$\,\AA,
$W_{\,\rm [O\,II]} > -5$\,\AA) and obtain $129$ passive galaxy spectra at $1 < z < 1.4$.  In both these
cases, the galaxies selected are typically of high stellar mass ($M_* > 10^{10}\rm\,M_{\odot}$; $> 97$ per
cent; see Fig.~\ref{stellar mass}).  Various properties of our high-$z$ PSB and passive spectra are shown in
Table.~\ref{sample properties}.  Note that for our passive selection, the addition of a $D_{4000}$ condition
to the criteria (i.e.\ to select older, more secure passive galaxies; see Section~\ref{Outflows}), has no
significant effect on the sample's median properties or the results of this work.

\begin{table}
\centering
\begin{minipage}{85mm}
\centering
\caption{\label{sample properties} Various properties of our high redshift ($1 < z < 1.4$) PSB and passive
galaxy spectra.  Median properties are shown for samples defined using either i)~spectroscopic criteria
(Spec), or ii)~a combination of both spectroscopic criteria and photometric PCA class
(Spec+PCA).}
\begin{tabular}{lccccc}
\hline
{Galaxy}				&\multicolumn{2}{c}{PSB}			&{}	&\multicolumn{2}{c}{Passive}			\\
{Property}				&{Spec}			&{Spec+PCA}		&{}	&{Spec}			&{Spec+PCA}		\\
\hline
{$N_{\rm spectra}$}			&{$41$}			&{$14$}			&{}	&{$129$}		&{$99$}			\\[1ex]
{-- $N$({\sc uds$_{z}$--fors$_{2}$})}	&{$25$}			&{$6$}			&{}	&{$63$}			&{$47$}			\\
{-- $N$({\sc uds$_{z}$--vimos})}	&{$6$}			&{$4$}			&{}	&{$7$}			&{$5$}			\\
{-- $N$({\sc m$_{16}$)}}		&{$7$}			&{$4$}			&{}	&{$12$}			&{$11$}			\\
{-- $N$({\sc vandels})}			&{$3$}			&{$0$}			&{}	&{$47$}			&{$36$}			\\
{}					&\multicolumn{5}{c}{----------------- Median values -----------------}					\\[1ex]
{$W_{\,\rm H\delta}$\rm\ (\AA)}		&{$7.24$}		&{$7.51$}		&{}	&{$1.44$}		&{$1.44$}		\\
{$W_{\,\rm [O\,II]}$\rm\ (\AA)}		&{$0.09$}		&{$-0.30$}		&{}	&{$1.83$}		&{$2.05$}		\\
{$D_{4000}$}				&{$1.24$}		&{$1.11$}		&{}	&{$1.34$}		&{$1.37$}		\\
{$z_{\rm spec}$	}			&{$1.19$}		&{$1.18$}		&{}	&{$1.13$}		&{$1.10$}		\\
{${\rm log_{10}}\,M_*/{\rm M_{\odot}}$}	&{$10.74$}		&{$10.67$}		&{}	&{$10.81$}		&{$10.84$}		\\
{$K_{\rm AB}$}				&{$20.89$}		&{$20.79$}		&{}	&{$20.67$}		&{$20.59$}		\\
{$r_{\rm e}$\ (kpc; $K_{\rm\,band}$)}	&{$2.08$}		&{$1.19$}		&{}	&{$1.96$}		&{$1.96$}		\\
{$n$\ ($K_{\rm\,band}$)}		&{$3.24$}		&{$3.88$}		&{}	&{$2.63$}		&{$2.84$}		\\
$\rm S/N^*$				&{$9.15$}		&{$12.29$}		&{}	&{$9.86$}		&{$10.06$}		\\
\hline
\end{tabular}
\end{minipage}
\begin{minipage}{80mm}
$^*$This is the median S/N of the individual spectra, as determined per resolution element and across the
observed spectral range.
\end{minipage}
\end{table}

In this study, we mainly focus on spectroscopically classified galaxy populations.  However, in the UDS field
robust galaxy classifications (i.e.\ passive, star-forming, PSB) are also available from the photometric
{\em `supercolour'} technique of \cite{Wild_etal:2014, Wild_etal:2016}, which is based on a Principal
Component Analysis~(PCA) of galaxy SEDs.  The effectiveness of this photometric selection, and in particular
its PSB classification, has recently been confirmed using the spectroscopic follow-up of
\cite{Maltby_etal:2016}.  In that paper, it was reported that $\sim80$ per cent of the photometrically
selected PSBs show the expected strong Balmer absorption (i.e.\ $W_{\,\rm H\delta} > 5$\,\AA) and that the
confirmation rate remains high ($\sim60$ per cent), even when stricter criteria are used to exclude cases
with significant [O\,{\sc ii}] emission.  We confirm that these findings hold for the extended spectral
samples ($z > 1$) used throughout this paper.  These PCA classifications use a much wider baseline in
wavelength than covered by our optical spectra (i.e.\ full SED information).  Consequently, the supercolour
technique is able to explicitly identify systems with a `hump' in the SED around the Balmer region
($\lambda\sim3500$--$6500$\,\AA), which is characteristic of a dominant A/F star component.  In this study,
the addition of supercolour (PCA) class to our classification criteria will likely result in a sub-population
of PSBs that host a more dominant A/F star population, and therefore experienced a more significant
starburst.  We make use of this sub-sample (Spec+PCA; $14$~PSBs) in a discussion of our results in
Section~\ref{outflow origins}, and various relevant properties are shown in Table~\ref{sample properties}.
The properties of a photometrically selected sub-sample for passive galaxies are also shown for completeness,
but not used in this work.

\subsection[]{Generating stacked spectra}

\label{Composite spectra}

\begin{figure*}
\includegraphics[width=0.99\textwidth]{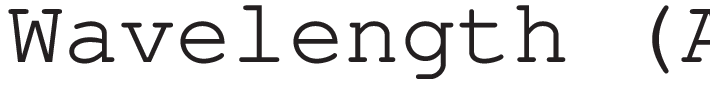}
\centering
\vspace{0.10cm}
\caption{\label{red stack} Red-optimised stacks: stacked optical spectra for our high redshift
($1 < z < 1.4$) PSB and passive galaxies, as determined from spectroscopic criteria (left-hand and
right-hand panels, respectively).  For these stacks, the individual rest-frame spectra were combined
following a flux normalisation over the Balmer break region ($3800 < \lambda_{\rm rest} < 4170$\,\AA).
Relevant sample sizes are shown in the legend, along with various spectral measurements
(e.g.\ $W_{\rm\,H\delta}$, $W_{\rm\,[O\,II]}$, $D_{4000}$, $\sigma_*$).  Uncertainties in these spectral
measurements ($1\sigma$) are determined from the variance between measurements performed on $1000$ simulated
spectra generated via a bootstrap method.  The errors in the stacked spectra (grey-shaded region) are
$1\sigma$ confidence limits (see Section~\ref{Composite spectra}).  In each case, a full spectral fit
obtained from {\sc ppxf} is shown for both the stellar component (red line) and the gas emission lines (cyan
line).
\vspace{0.10cm}}
\end{figure*}

In order to determine the presence of gaseous outflows, we use the Mg\,{\sc ii} absorption doublet
($\lambda\lambda\,2796$,\,$2803$\,\AA), which is a sensitive tracer of low-ionisation interstellar gas
($T\sim10^4\rm\,K$).  We note that systemic-frame Mg\,{\sc ii} absorption can originate from either the ISM
or stellar photospheres, but the detection of a blue-shifted component to this absorption feature is
generally thought to indicate galactic-scale outflows along the line-of-sight to the observer.  In this
study, $\sim65$ per cent of our spectroscopically-classified galaxies (see Table~\ref{sample properties})
have full coverage of the Mg\,{\sc ii} region ($26/41$~PSB and $84/129$ passive spectra).  Unfortunately, the
signal-to-noise (S/N) of these VIMOS/FORS2 spectra is not sufficient to reliably determine the structure of
the Mg\,{\sc ii} profile on an individual galaxy basis [typically
$\rm S/N(\lambda_{\rm rest}\sim2800$\,\AA$)\sim5$].\footnote{Note: throughout this study, S/N is defined per
resolution element.}  We therefore increase the effective S/N via a stacking analysis, combining the
individual rest-frame spectra following an optimised flux normalisation.  The following procedure is used.
\begin{enumerate}
\item The individual spectra are shifted to their respective rest-frame and oversampled onto a common and
finer dispersion axis ($\Delta\lambda = 0.25$\,\AA).  For this we use the spectroscopic redshifts
($z_{\rm spec}$) computed with {\sc ez}, which are determined using a cross-correlation of spectral templates
(see Section~\ref{The UDS}).  We note that using an alternative $z_{\rm spec}$, defined using a single
stellar absorption feature which is present in all our spectra, i.e.\ Ca\,{\sc ii K} ($\lambda3933.7$\,\AA),
has no significant effect on the results of this work.
\vspace{0.10cm}
\item We combine the individual rest-frame spectra following an optimised flux normalisation.  In this study,
in addition to our analysis of the Mg\,{\sc ii} absorption feature ($\lambda_{\rm rest}\sim2800$\,\AA), we
also wish to obtain stellar velocity dispersions $\sigma_*$ using features at
$\lambda_{\rm rest} > 3550$\,\AA\ (see Section~\ref{velocity dispersions}).  Therefore, to optimise our
analysis we generate two median-stacked spectra: (a)~a red-optimised stack
($\lambda_{\rm rest} > 3550$\,\AA), using the full spectroscopically classified sample (see
Table~\ref{sample properties}) and a flux normalisation over the Balmer break region
($3800 < \lambda_{\rm rest} < 4170$\,\AA); and (b)~a blue-optimised stack
($\lambda_{\rm rest}\sim2800$\,\AA), using only spectra with Mg\,{\sc ii} coverage and a flux normalisation
over the Mg\,{\sc ii} continuum ($2700 < \lambda_{\rm rest} < 2900$\,\AA).  In both cases, the normalised
spectra are averaged without any weighting to avoid any bias towards the brightest galaxies with the highest
S/N.  For the blue-optimised stack, we also apply a nominal S/N cut to the individual spectra, as determined
over the Mg\,{\sc ii} region [$\rm S/N(\lambda_{\rm rest}\sim2800$\,\AA$) > 1.5$], in order to remove the
influence of poor quality spectra on our stacked Mg\,{\sc ii} profiles ($< 10$~per~cent of our high-$z$
spectra).
\end{enumerate}
The final red-optimised stacks for both our high-$z$ PSB and passive galaxies are shown in
Fig.~\ref{red stack}.  These spectra are used to determine the typical stellar velocity dispersions
$\sigma_*$ of their respective galaxy populations in Section~\ref{velocity dispersions}.  The blue-optimised
stacks are used to determine the presence of outflows in these galaxies and are presented in
Section~\ref{Outflows} (see Fig.~\ref{blue stack}).  The effective spectral resolution of these stacked
spectra $\Delta\lambda_{\rm\,FWHM}$ is $\sim5.8$\,\AA\ ($\sim180\pm12\rm\,km\,s^{-1}$).  Uncertainties in
these spectra are determined from the mean of the standard errors from $100$ simulated median-stacks
generated via a bootstrap technique.

In this study, we note that while the majority of our individual spectra have a similar spectral resolution
($R\sim600$), our sample does include a small number of low-resolution spectra ($R\sim200$) from
UDSz-{\sc vimos} (see Table~\ref{sample properties}).  These low-resolution spectra are of high S/N and
include some of the brightest and most significant PSB spectra within our sample. Consequently, in order to
maximise the effective S/N, and prevent biasing our sample against these galaxies, we include these
low-resolution spectra in our stacking analysis.  However, we note that removing these spectra from our
samples has no major impact on our Mg\,{\sc ii} analysis or conclusions.  Furthermore, a consistent
Mg\,{\sc ii} profile and PSB outflow velocity is obtained using just these low-resolution spectra.  We also
obtain consistent results if all spectra are reduced in resolution to~$R\sim200$.

\vspace{-0.2cm}

\section[]{Results and discussion}

\label{Results}

\subsection[]{Stellar velocity dispersion $\sigma_*$}

\label{velocity dispersions}

To gain insight into our high-$z$ PSB and passive galaxies, we perform full spectral fitting on their
stacked spectra (red-optimised;~Fig.~\ref{red stack}) using the penalized pixel-fitting method ({\sc ppxf};
\citealt{Cappellari&Emsellem:2004, Cappellari:2017}) and the MILES stellar templates
\citep{Vazdekis_etal:2010}.  These fits can be used to determine the typical stellar velocity dispersion
$\sigma_*$ of our galaxy populations, provided a suitable estimate for any additional sources of spectral
broadening.  For example, the effective instrumental response ($\sigma_{\rm instr}$) and the broadening
introduced by stacking ($\sigma_{\rm stack}$).  In this study, we assume all broadening functions are
Gaussian, consequently $\sigma_*$ can be recovered from the observed dispersion $\sigma_{\rm obs}$ following
\cite{Cappellari_etal:2009},
\begin{equation}
\sigma_* = \sqrt{\sigma_{\rm obs}^2 - \sigma_{\rm instr}^2 - \sigma_{\rm stack}^2}.
\end{equation}
However, to determine $\sigma_*$, {\sc ppxf} requires the stellar templates used for spectral fitting to
have a spectral resolution $\Delta\lambda$ that matches that of the stacked spectrum.  The additional
broadening required for the best fit is then used to determine $\sigma_*$.  In this study, we use the MILES
stellar library, which is an empirical library of stellar templates covering the optical regime
($\lambda\,3525$--$7500$\,\AA).  These templates have a well-defined spectral resolution of
$\Delta\lambda_{\rm\,FWHM} = 2.51$\,\AA\ \citep{Sanchez-Blazquez_etal:2006, Falcon-Barroso_etal:2011}, which
is very different to that of our stacked spectra ($\Delta\lambda_{\rm\,FWHM} \sim 5.8$\,\AA; see
Section~\ref{Composite spectra}).  Consequently, prior to fitting, we broaden these stellar templates to
match that of the effective spectral resolution of our stacked spectra $\Delta\lambda_{\rm eff}$ (FWHM),
where
\begin{equation}
\Delta\lambda_{\rm eff} =  2.355\times\sqrt{\sigma_{\rm instr}^2+\sigma_{\rm stack}^2}.
\end{equation}

For each stacked spectrum, we estimate the effective $\sigma_{\rm instr}$ using the
median\,[$\sigma^{*}_{\rm instr}/(1+z)$] of the input spectra, where $\sigma^{*}_{\rm instr}$ is the
observed-frame instrument response as determined from the resolving power $R$ of the respective spectrograph
(see Section~\ref{The UDS})\footnote{Note: in our $\sigma_*$ analysis, we retain the small fraction of
low-resolution UDSz--{\sc vimos} spectra ($R\sim200$) that contribute to our final stacked spectra (see
Table~\ref{sample properties}).  However, we note that removing these spectra from our analysis has no
significant impact on the $\sigma_*$ measurements for our galaxy populations, or our conclusions.}.  In this
study, we find $\sigma_{\rm instr}$ is typically $\sim2.45\pm0.16$\,\AA\ ($\sim180\pm12\rm\,km\,s^{-1}$).
The broadening related to stacking spectra, $\sigma_{\rm stack}$, originates from redshift errors $\Delta{z}$
and the error $\Delta\lambda$ introduced by shifting the individual spectra to their rest-frame.  In this
study, we use an indicative value for $\sigma_{\rm stack}$ estimated using the variance in $\Delta\lambda$,
obtained from the redshift measurements of $1000$ simulated spectra with the same spectral resolution,
wavelength sampling and $S/N$ as our observations.  From this we determine that
$\sigma_{\rm\,stack} < 0.53$\,\AA\ ($<40\rm\,km\,s^{-1}$) and is essentially negligible with respect to the
instrumental and intrinsic broadening.  Finally, to determine the typical $\sigma_*$ from our stacked
spectra, we perform spectral fits with {\sc ppxf} over the wavelength range $3550$--$4550$\,\AA, and using a
model consisting of the stellar component plus the [O\,{\sc ii}] and H$\delta$ emission lines.  The final
$\sigma_*$ and its respective $1\sigma$ uncertainty are determined using the median and variance of fits
performed on $1000$ simulated spectra generated via a bootstrap analysis.

\newpage

\begin{figure}
\includegraphics[width=0.49\textwidth]{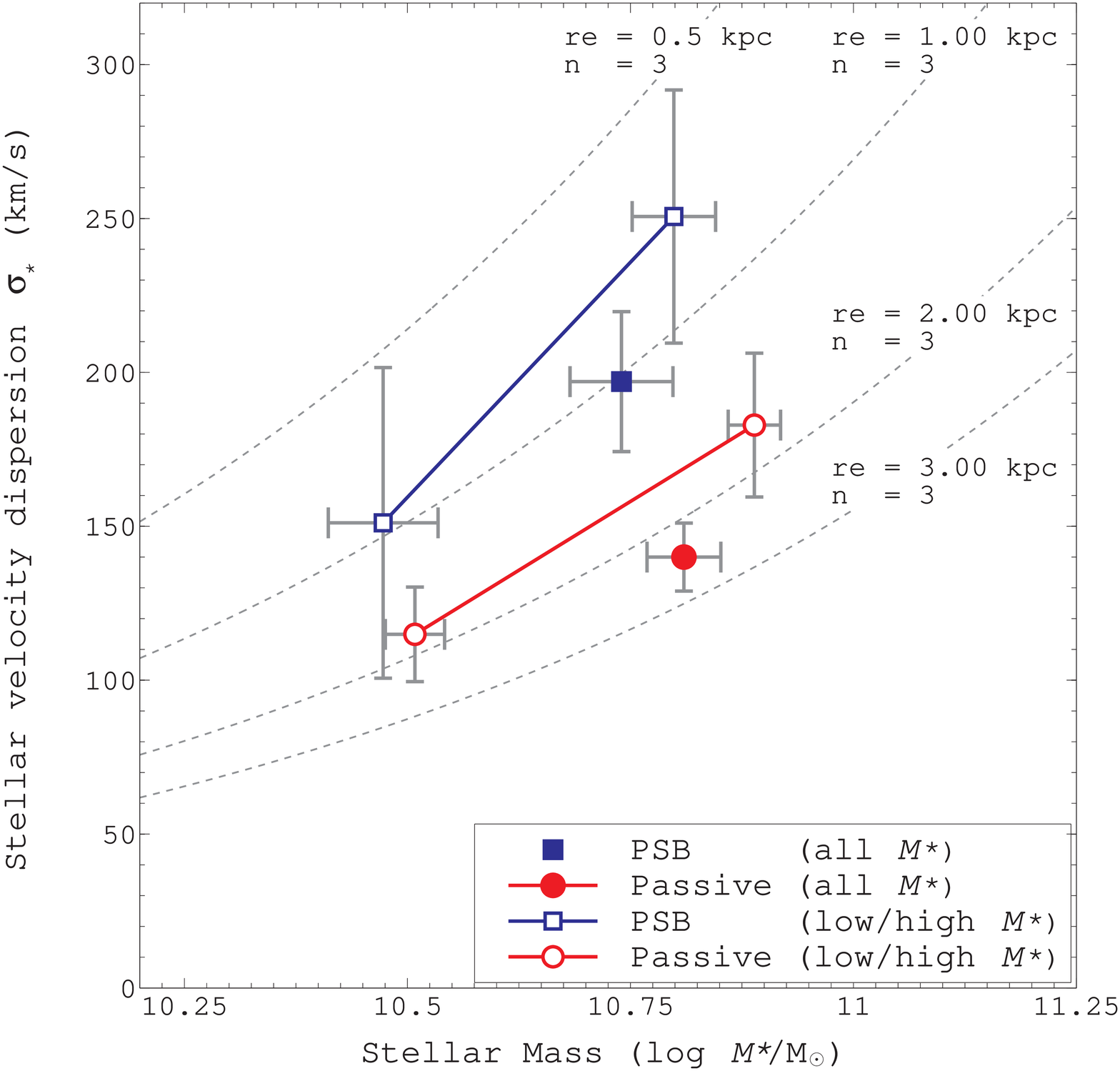}\\
\centering
\vspace{-0.2cm}
\caption{\label{velocity dispersion} Stellar velocity dispersion $\sigma_*$ as a function of stellar
mass~$M_*$ for our high-$z$ spectroscopically classified PSB and passive galaxies.  Results are shown for
both our full sample (filled symbols), and that separated by $M_*$ (open symbols).  In the latter, we
separate our sample into  low mass (${\rm log_{10}}\,M_*/{\rm M_{\odot}} < 10.7$) and high mass
(${\rm log_{10}}\,M_*/{\rm M_{\odot}} > 10.7$).  The $\sigma_*$ measurements are plotted at the median $M_*$
of the respective sample, with $1\sigma$ uncertainties determined from the variance between measurements
performed on $1000$ simulated spectra generated via a bootstrap method.  Uncertainties in $M_*$ ($1\sigma$)
are the standard errors in the median for the respective sample.  As expected, $\sigma_*$ increases with
$M_*$ for both populations.  For high-$z$ PSBs, there is a tentative indication that they present a higher
$\sigma_*$ than analogous passive galaxies, particularly at the highest masses
($M_* > 10^{10.7}\rm\,M_{\odot}$).  In this figure, we compare these results to the dynamical mass
$M_{\rm d}$($\sigma_*$,~$r_{\rm e}$,~$n$) relations, for various structural configurations (see
Equation~\ref{dynamical mass equation}; dashed lines).
\vspace{-0.1cm}}
\end{figure}

For our high-$z$ PSB and passive galaxies, the resultant {\sc ppxf} fits to their stacked spectra are shown
in Fig.~\ref{red stack}.  In all cases, our stacked spectra are well-modelled by the resultant spectral
fits, with reduced chi-squared values of $\chi^2_{\rm red}\sim1$.  These fits yield high $\sigma_*$ values
for both our galaxy populations [$\sigma_{*}(\rm\,PSB)\sim200\pm23\rm\,km\,s^{-1}$;
$\sigma_{*}(\rm\,passive)\sim140\pm11\,\rm km\,s^{-1}$], as expected for their high stellar masses (median
$M_* \sim 10^{10.7}\rm\,M_{\odot}$; see Table~\ref{sample properties} and Fig.~\ref{velocity dispersion}).

\begin{table}
\centering
\begin{minipage}{85mm}
\centering
\caption{\label{vdisp-vs-M}The typical stellar velocity dispersion $\sigma_*$, structural parameters
($r_{\rm e}$,~$n$) and kinematically derived dynamical masses $M_{\rm d}$ for our high-$z$ PSB and passive
galaxy spectra (Spec).  Results are also shown for the Spec+PCA sub-samples (see
Section~\ref{Sample selection}).}
\begin{tabular}{lccccc}
\hline
{Galaxy}					&\multicolumn{2}{c}{PSB}			&{}	&\multicolumn{2}{c}{Passive}			\\
{Property}					&{Spec}			&{Spec+PCA}		&{}	&{Spec}			&{Spec+PCA}		\\
\hline
{$\sigma_* \rm\ (km\,s^{-1})$}			&{$197\pm23$}		&{$255\pm33$}		&{}	&{$140\pm11$}		&{$149\pm13$}		\\
{$r_{\rm e}$\,(kpc;\,$K_{\rm\,band}$)}		&{$2.08$}		&{$1.19$}		&{}	&{$1.96$}		&{$1.96$}		\\
{$n$ ($K_{\rm\,band}$)}				&{$3.24$}		&{$3.88$}		&{}	&{$2.63$}		&{$2.84$}		\\
{${\rm log_{10}}\,M_{\rm d}/{\rm M_{\odot}}^*$}	&{$11.02$}		&{$10.94$}		&{}	&{$10.76$}		&{$10.79$}		\\
\hline
\multicolumn{6}{l}{*Derived using Equation~\ref{dynamical mass equation}.}\\
\end{tabular}
\end{minipage}
\end{table}

From the Scalar Virial Theorem, it is well established that $\sigma_*$ is related to dynamical mass
$M_{\rm d}$,
\begin{equation}
\label{dynamical mass equation}
M_{\rm d} \approx k_{\rm d}\,\frac{\sigma_*^2r_{\rm e}}{G},
\end{equation}
where $G$ is the gravitational constant and $k_{\rm d}$ is the virial coefficient.  This coefficient
($k_{\rm d}$) takes into account projection effects and the structure of the mass distribution.  Under the
assumption that the mass follows a S{\'e}rsic distribution, the virial coefficient~$k_{\rm d}$ has been
computed by several authors \citep[e.g.][]{Prugniel&Simien:1997, Bertin_etal:2002, Cappellari_etal:2006}.
For example, \cite{Bertin_etal:2002} provide a simple analytical approximation,
\begin{equation}
k_{\rm d}(n) \approx \frac{73.32}{10.465 + (n - 0.94)^2} + 0.954,
\end{equation}
(see also \citealt{Taylor_etal:2010} and \citealt{Zahid&Geller:2017}, for useful explanations).  Therefore,
using the $K$-band structural parameters for our galaxy populations (see Table~\ref{sample properties}), we
can use our $\sigma_*$ measurements to estimate their typical dynamical mass $M_{\rm d}$.  These estimates
confirm that both our PSB and passive galaxies are intrinsically massive in nature
($M_{\rm d}\sim10^{11}\rm\,M_{\odot}$; see Table~\ref{vdisp-vs-M}).  We note that consistent results are
obtained using alternative derivations for the dynamical mass \citep[e.g.][who find that $M_{\rm d}$ is
approximately twice the total mass within $r_{\rm e}$]{Cappellari_etal:2006, Cappellari_etal:2013}.  Finally,
we note that the virial coefficient $k_{\rm d}$ assumed in this work \citep{Bertin_etal:2002} is derived
assuming a central $\sigma_*$ measurement (aperture~$r < \frac{1}{8}\,r_{\rm e}$), while our $\sigma_*$
measurements are likely to be averaged over a much larger aperture.  However, based on the aperture
$\sigma_*$ corrections from previous studies \citep[e.g.][]{Cappellari_etal:2006}, the difference between our
$\sigma_*$ measurements and the central $\sigma_*$ is expected to be $\lesssim20$ per cent.  Therefore, this
issue is not expected to have a significant impact on the results of this work.

\begin{table}
\centering
\begin{minipage}{85mm}
\centering
\caption{\label{velocity dispersions by mass}The typical stellar velocity dispersion $\sigma_*$ for our
high-$z$ spectroscopically classified PSB and passive galaxy spectra, in different stellar mass $M_*$ ranges.
We separate our sample into low $M_*$ (${\rm log_{10}}\,M_*/{\rm M_{\odot}} < 10.7$) and high $M_*$
(${\rm log_{10}}\,M_*/{\rm M_{\odot}} > 10.7$ ).  The stellar masses presented in this table, are the median
$M_*$ in each mass range.  Typical structural parameters ($r_{\rm e}$, $n$) and dynamical mass $M_{\rm d}$
estimates are also presented.}
\begin{tabular}{lccccc}
\hline
{Galaxy}					&\multicolumn{2}{c}{PSB}		&{}	&\multicolumn{2}{c}{Passive}		\\
{Property}					&{Low $M_*$}		&{High $M_*$}	&{}	&{Low $M_*$}		&{High $M_*$}	\\
\hline
{$N_{\rm spectra}$}				&{$18$}			&{$23$}		&{}	&{$55$}			&{$74$}		\\
{$\sigma_* \rm\ (km\,s^{-1})$}			&{$151\pm50$}		&{$251\pm41$}	&{}	&{$115\pm15$}		&{$183\pm23$}	\\
{$r_{\rm e}$ (kpc;\,$K_{\rm\,band}$)}		&{$1.20$}		&{$2.27$}	&{}	&{$1.37$}		&{$2.40$}	\\
{$n$ ($K_{\rm\,band}$)}				&{$3.48$}		&{$3.08$}	&{}	&{$2.45$}		&{$2.78$}	\\
{${\rm log_{10}}\,M_*/{\rm M_{\odot}}$}		&{$10.47$}		&{$10.89$}	&{}	&{$10.50$}		&{$10.96$}	\\
{${\rm log_{10}}\,M_{\rm d}/{\rm M_{\odot}}^*$} &{$10.52$}  		&{$11.29$} 	&{}	&{$10.45$}		&{$11.07$}	\\
\hline
\multicolumn{6}{l}{*Derived using Equation~\ref{dynamical mass equation}.}\\
\end{tabular}
\end{minipage}
\end{table}

Finally, to expand on our $\sigma_*$ results, we also separate our galaxy populations by stellar mass $M_*$,
and obtain the typical $\sigma_*$ measurements from the resultant stacked spectra (for the $M_*$ distribution
of our samples, see Fig.~\ref{stellar mass}).  We separate our sample into low mass
(${\rm log_{10}}\,M_*/{\rm M_{\odot}} < 10.7$) and high mass (${\rm log_{10}}\,M_*/{\rm M_{\odot}} > 10.7$),
and compare the resultant $\sigma_*$ with the median $M_*$ in each sub-sample (see
Fig.~\ref{velocity dispersion} and Table~\ref{velocity dispersions by mass}).  As expected, we find that
$\sigma_*$ increases with $M_*$ for both the PSB and passive galaxy populations.  Furthermore, we find that
for high-$z$ PSBs, there is a tentative indication that they present higher $\sigma_*$ than analogous
passive galaxies, particularly at the highest masses ($M_* > 10^{10.7}\rm\,M_{\odot}$).  We compare these
results to the dynamical mass $M_{\rm d}$($\sigma_*$, $r_{\rm e}$, $n$) relations from
Equation~\ref{dynamical mass equation}, for various structural configurations (see
Fig.~\ref{velocity dispersion}).  Although these relations are only applicable for dynamical mass
$M_{\rm d}$, they suggest that our $\sigma_*$--$M_*$ results indicate that high-$z$ PSBs are slightly more
compact (i.e.\ smaller $r_{\rm e}$) than passive galaxies, at the same stellar mass.  This is consistent with
the findings of \cite{Almaini_etal:2017} and \cite{Maltby_etal:2018} for this galaxy population, who used the
photometric PCA (i.e.\ {\em supercolour}) classifications and galaxy structural
parameters~\mbox{($r_{\rm e}$, $n$)}.  Overall, we suggest that these results are consistent with a recent
compaction event for high-$z$ PSBs, which may have triggered the preceding starburst, potentially
high-velocity outflows (see Section~\ref{Outflows}) and subsequent quenching.

\subsection[]{Measuring outflows from Mg\,{\sc ii} absorption}

\label{Outflows}

\begin{figure*}
\includegraphics[width=0.99\textwidth]{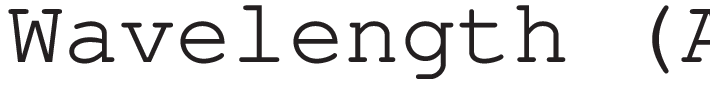}
\centering
\vspace{0.10cm}
\caption{\label{blue stack} Blue-optimised stacks: stacked optical spectra for our high redshift
($1 < z < 1.4$) PSB and passive galaxies, as determined from spectroscopic criteria (left-hand and
right-hand panels, respectively).  For these stacks, the individual rest-frame spectra were combined
following a flux normalisation over the Mg\,{\sc ii} continuum ($2700 < \lambda_{\rm rest} < 2900$\,\AA).
Relevant sample sizes are shown in the legend, along with various spectral measurements
(e.g.\ $W_{\rm\,H\delta}$, $W_{\rm\,[O\,II]}$, $D_{4000}$).  Uncertainties in these spectral measurements
($1\sigma$) are determined from the variance between measurements performed on $1000$ simulated spectra
generated via a bootstrap method.  The errors in the stacked spectra (grey-shaded region) are $1\sigma$
confidence limits (see Section~\ref{Composite spectra}).  For each spectrum, the sub-panel shows the best fit
to the Mg\,{\sc ii} absorption profile, using a model comprising two Gaussian-convolved doublets: one fixed
at the systemic redshift (red line), and another with a free centroid to model the outflow (green line).  The
outflow component is only included in the model if required (as determined by an F-test).  The rest-frame
wavelengths of the Mg\,{\sc ii} doublet are also shown for reference (red-dashed lines).  We find evidence
for high-velocity outflows ($\Delta{v}\sim1500\pm150\rm\,km\,s^{-1}$) in our high-$z$ PSBs, potentially
representing the residual signature of a feedback process which quenched these galaxies.  In contrast, we
find no significant evidence for outflows in our high-$z$ passive galaxies (the population our high-$z$ PSBs
will most likely evolve into).
\vspace{0.10cm}}
\end{figure*}

In this section, we determine the prevalence of galaxy-scale outflows in our galaxy populations using their
blue-optimised stacked spectra (see Section~\ref{Composite spectra} and Fig.~\ref{blue stack}).  To achieve
this, we analyse the structure of the Mg\,{\sc ii} absorption profile, which is a sensitive tracer of the
low-ionisation interstellar medium (ISM). For our high-$z$ PSBs, an initial inspection of their stacked
spectrum reveals significant asymmetry in the Mg\,{\sc ii} profile (see Fig.~\ref{blue stack}).  With respect
to the central systemic wavelength of the Mg\,{\sc ii} doublet ($\lambda2799.5$\,\AA), there is a clear
excess of absorption towards bluer wavelengths.  In contrast, for high-$z$ passive galaxies, no such
asymmetry or excess blue absorption is observed.

To determine the significance of this result, and detect the presence of any outflowing component, we use the
following procedure.  First, we normalise for the continuum across Mg\,{\sc ii}, using a smoothing spline fit
to the continuum flux on either side of the Mg\,{\sc ii} feature.  We then model the Mg\,{\sc ii} absorption
profile using either one or two components, as follows.
\begin{enumerate}
\item{\it One component (systemic absorption):}  the Mg\,{\sc ii} absorption is modelled using a single
component, fixed at the rest-frame wavelength for the systemic absorption (ISM + stellar).  This component
consists of a doublet ($\lambda\lambda\,2796$,\,$2803$\,\AA), with an intensity ratio of 1.2:1, as observed
for massive high-$z$ galaxies in the high resolution spectra of \cite{Weiner_etal:2009}.  In the fitting
process, each line is given an initial narrow width ($\sigma = 1$\AA), and then convolved with a Gaussian to
model the spectral broadening, which is necessary since the Mg\,{\sc ii} doublet is essentially unresolved in
our spectra.  In each case, the width of the Gaussian used for convolution is fixed using an initial fit to
only the red-side of the Mg\,{\sc ii} profile (i.e.~$\lambda_{\rm rest} > 2803$\,\AA).  This region of the
absorption profile is largely unaffected by the potential contamination from any outflowing component, and
therefore provides a suitable estimate for the intrinsic width of the absorption line.  For our high-$z$
PSBs, this model reveals that $\sim25\pm5$ per cent of the absorption on the blue-side of the Mg\,{\sc ii}
profile ($\lambda_{\rm rest} < 2799.5\,$\AA) is  not accounted for by the systemic component.
\vspace{0.10cm}
\item{\it Two components (systemic absorption + outflow):} the Mg\,{\sc ii} absorption is modelled using two
components, one fixed at the rest-frame wavelength for the systemic absorption (ISM + stellar), and another
with a free centroid to model the outflow.  Each component consists of a Gaussian-convolved doublet, as
described above.\footnote{Note: with this two-component model, we find consistent results are also obtained
when using a free-width Gaussian for the convolution.}  This simple model yields an offset $\Delta\lambda$ of
the outflowing component with respect to the systemic-frame wavelength, which can be used to determine its
characteristic velocity offset $\Delta{v}$ from the systemic redshift.\footnote{Note: here the systemic
redshift is defined as the template-fitting $z_{\rm spec}$ from {\sc ez}.  However, entirely consistent
offsets $\Delta{v}$ are also obtained if a strong stellar absorption line (e.g.\ Ca\,{\sc ii K}) is used to
define the systemic redshift.}  For our stacked spectra, $\Delta{v}$ represents an estimate of the typical
outflow velocity in the low-ionisation gas for our galaxy populations.  However, based on simulations, we
note that $\Delta{v}$ will likely be an over-estimate of the actual median outflow velocity $v_{\rm out}$ by
$\sim350\rm\,km\,s^{-1}$ (see Appendix~\ref{appendix A}).  The $1\sigma$ uncertainties in these velocity
measurements are determined using the variance between analogous fits performed on $1000$ simulated spectra
generated via a bootstrap analysis.
\end{enumerate}
To determine which of these models best describes the Mg\,{\sc ii} profile, and therefore determine the
presence of an outflowing component, we use an F-test.  Formally, the two-component model will always provide
the best fit to the data, but an F-test can be used to determine whether the additional outflowing component
is statistically required.  This F-test yields a $p$-value for accepting the null hypothesis (i.e.\ that an
outflowing component is not required), and rejects the two-component model if $p > 0.05$.

For our high-$z$ PSB and passive galaxies, the relevant fits to the Mg\,{\sc ii} profile are presented in
Fig.~\ref{blue stack}.  For high-$z$ PSBs, the Mg\,{\sc ii} profile presents a significant excess of
blue-shifted absorption ($\sim25$~per~cent).  In this case, our best-fitting two-component model yields an
outflow component with a large velocity offset ($\Delta{v}\sim1500\pm150\rm\,km\,s^{-1}$), indicating these
galaxies host high-velocity outflows in their interstellar medium (ISM).  Based on our simulations, this
$\Delta{v}$ corresponds to a typical outflow velocity of $v_{\rm out}\sim1150\pm160\rm\,km\,s^{-1}$ (see
Appendix~\ref{appendix A}).  The significance of this outflowing component is $>3\sigma$, as determined by an
F-test (\,$p < 0.003$).  In contrast, for high-$z$ passive galaxies, we find that no significant outflow
component is required to account for their Mg\,{\sc ii} profile, which is also confirmed by an F-test
(\,$p > 0.05$).  Furthermore, the weaker Mg\,{\sc i} absorption line, which is another tracer of the
low-ionisation ISM, also shows no signs of a significant outflowing component.

For our high-$z$ PSBs, the outflow velocities we measure depend significantly on the correct modelling of the
systemic component.  However, an alternative {\em boxcar} method can also be used to measure outflow
velocities, and this does not suffer from this dependency \citep[see e.g.][]{Rubin_etal:2010,
Bordoloi_etal:2014}.  In this method, the mean outflow velocity $\langle v_{\rm out} \rangle$ is estimated
from the global shift of the observed absorption line as
follows,
\begin{equation}
\langle v_{\rm out} \rangle = \frac{W_{\rm total}}{W_{\rm out}} \langle v_{\rm total} \rangle.
\end{equation}
Here $W_{\rm total}$ and $W_{\rm out}$ are the equivalent widths of the full Mg\,{\sc ii} absorption profile
and its outflowing component, respectively, and $\langle v_{\rm total} \rangle$ is the mean absorption
weighted velocity of the observed absorption line.  To determine $W_{\rm out}$, the difference in equivalent
width between the red and blue-side of the Mg\,{\sc ii} absorption profile is used (see
\citealt{Rubin_etal:2010} and \citealt{Bordoloi_etal:2014}, for further details).  For our high-$z$ PSBs, we
use this method to confirm the presence of high-velocity outflows, obtaining a mean outflow velocity of
\mbox{$809\pm104\rm\,km\,s^{-1}$} for this population.  The $1\sigma$ uncertainty in this velocity is
determined using the variance between measurements performed on $1000$ simulated spectra generated via a
bootstrap analysis.  We note that this outflow velocity is slightly lower than that obtained from our
decomposition method, but nonetheless confirms the presence of high-velocity outflows in our high-$z$ PSBs.

With respect to the outflowing component, the absorption strength (i.e.\ equivalent width) can also provide
insight into the nature of outflows in our high-$z$ PSBs.  Since Mg\,{\sc ii} absorbing gas is optically
thick at low column densities {($N_{\rm\,Mg\,II}\gtrsim10^{13}\rm\,atoms\,cm^{-2}$), if present, the
absorption is generally saturated in both the stellar and ISM components
\citep[see e.g.][]{Weiner_etal:2009}.  This is apparent in our high-$z$ passive galaxies, where the
Mg\,{\sc ii} doublet is just resolved in their stacked profile (see Fig.~\ref{blue stack}).  For
non-saturated Mg\,{\sc ii} absorption the line ratio is 2:1, but in this case it is close to 1:1, indicating
near saturated absorption.  For our high-$z$ PSBs, we note that despite the expected saturation in both the
systemic and outflowing components, the Mg\,{\sc ii} profile is dominated by the systemic absorption (see
Fig.~\ref{blue stack}).  In a stacked spectrum, this could be due to a combination of i)~high-velocity
outflows being present in only a fraction~$D$ of the sample; and ii)~the typical covering fraction
$\langle C_{\rm f} \rangle$ of the outflowing wind, which is the fraction of the stellar distribution it
obscures along the line-of-sight (e.g.\ due to a collimated and/or clumpy outflow).  Consequently, for the
outflowing component, the Mg\,{\sc ii} absorption depth can be defined as
$A_{\rm d}= D\langle C_{\rm f} \rangle$ and is $A_{\rm d}\sim0.2$ for our high-$z$ PSBs (see
Fig.~\ref{blue stack}).  In comparison to the stacked spectra of previous works, we find this absorption
depth to be significantly lower than that of massive star-forming/starburst galaxies at $z\sim1.4$
\citep[where $A_{\rm d}\sim 0.55$;][]{Weiner_etal:2009}.  This difference could be explained by a lower
covering fraction $\langle C_{\rm f} \rangle$, but we suggest the more likely explanation is due to a lower
detection fraction $D$ in our PSBs, potentially due to the high-velocity outflows only persisting for the
early PSB phase (see Section~\ref{outflow origins}).  Based on the typical covering fraction of local
starbursts \citep[$C_{\rm f}$ = $0.4$--$0.5$;][]{Rupke_etal:2005}, we estimate that $D\sim0.5$ for our
high-$z$ PSB outflows.

With respect to the systemic component, since stellar Mg\,{\sc ii} absorption is known to increase in
strength for older stellar populations \cite[see e.g.][]{Martin&Bouche:2009}, for our PSB and passive spectra
the dominant contribution to the systemic Mg\,{\sc ii} is expected to be stellar in origin.  In this work, we
have modelled the systemic absorption (ISM + stellar) as a single component comprising a Gaussian-convolved
doublet.  An alternative approach is to use synthetic (i.e.~theoretical) stellar libraries
\citep[e.g.~UVBLUE,][]{Rodriguez_etal:2005, Coelho_etal:2014}, to estimate and remove the stellar
Mg\,{\sc ii} component using full spectral fitting.  However, we note that while these theoretical stellar
libraries cover the UV region ($\lambda\sim2800$\,\AA) they are not as robust as the empirical libraries
available in the optical regime \citep[e.g.\ MILES; ][]{Vazdekis_etal:2010}.  For example, it is known that
i)~various regions of the UV spectrum are poorly reproduced; and ii)~the prominent metallic lines in F/G
stars (including Mg\,{\sc ii}), are always stronger in the synthetic spectra than in observed stars
\citep[see][for further details]{Rodriguez_etal:2005}.  This is also true to a lesser extent for A-stars
\citep[see fig.~10 from ][]{Rodriguez_etal:2005}.  Due to these uncertainties, and since A/F stars will be a
significant component in our PSB spectra, we have chosen not to adopt this approach for our primary analysis
of the Mg\,{\sc ii} profile.  However, we have explored this issue in detail (see Appendix~\ref{appendix B}),
and confirm that consistent outflow velocities are obtained using this alternative approach.  Although, for
our high-$z$ PSBs, we also note that while the significance of the outflowing component remains $>3\sigma$,
the strength (i.e.\ absorption depth $A_{\rm d}$) is reduced.  This is consistent with either a smaller
covering fraction~$C_{\rm f}$ or lower detection fraction~$D$ for these winds than indicated by our
two-component Gaussian model (see Appendix~\ref{appendix B}).

\subsection[]{The origin of high-velocity outflows in PSBs}

\label{outflow origins}

To expand on our results, we use the $D_{4000}$ index, which is a proxy for both the mean age and metallicity
of a galaxy's stellar population \citep{Bruzual:1983}.  This index measures the strength of the $4000$\,\AA\
break, and is consequently small for young stellar populations, and larger for both older, and more
metal-rich galaxies.  In this study,
\begin{equation}
D_{4000} = \frac{\langle F_{\nu}\,(\lambda\,\text{4000--4100\,\AA})
\rangle}{\langle F_{\nu}\,(\lambda\,\text{3850--3950\,\AA}) \rangle},
\end{equation}
following the revised definition outlined by \cite{Balogh_etal:1999}.  From the individual spectra, the
median value of $D_{4000}$ is $\sim1.24$ for our PSBs, and $\sim1.34$ for our passive galaxies (see
Table~\ref{sample properties}).  Using the $D_{4000}$ index to divide our spectral samples, we find a
tentative hint that high-velocity outflows are more significant (as determined by an F-test; see
Section~\ref{Outflows}) in the stacked spectra of younger PSB galaxies (i.e.~$D_{4000} < 1.24$).  This
potential relationship between high-velocity outflows and star-formation history (SFH) will be explored in
more detail in future work.  For our passive galaxies, where no outflows are detected, we find that the
addition of a $D_{4000}$ condition to their selection criteria, e.g.~to select either younger
($D_{4000} < 1.4$), or older, more secure passive galaxies ($D_{4000} > 1.4$), has no significant effect on
the nature of their stacked Mg\,{\sc ii} profile or the lack of an outflowing component.  If confirmed, these
results might suggest that for our PSBs, the high-velocity outflows were launched during, or shortly after,
the preceding starburst, and may have ceased by the time the galaxy becomes truly passive.  This would
suggest an inherent relationship between the high-velocity wind and the quenching of star formation.

In comparison to previous works, we find that for high-$z$ PSBs the high-velocity outflows we detect
($v_{\rm out}\sim1150\rm\,km\,s^{-1}$) are much faster than those observed in typical star-forming galaxies
at this epoch \citep[e.g.][]{Talia_etal:2012, Bradshaw_etal:2013, Bordoloi_etal:2014}.  Such high-velocity
outflows are only consistent with those of either star-forming AGN \citep[e.g.][]{Hainline_etal:2011,
Harrison_etal:2012, Talia_etal:2017} or the massive \mbox($M_* > 10^{10.5}\rm\,M_{\odot}$)}
starburst/post-starburst galaxies observed at $z\sim0.6$ \citep[e.g.][]{Tremonti_etal:2007, Geach_etal:2014,
Sell_etal:2014}.  For our massive high-$z$ PSBs (median $M_*\sim10^{10.7}\rm\,M_{\odot}$; see
Table~\ref{sample properties}), this would be consistent with the high-velocity winds being launched during
the preceding starburst, rather than during a phase of more general star-forming activity.  In particular, it
is interesting that these high-$z$ outflows are consistent with those of the luminous, massive, young, but
also much rarer PSBs at lower redshift \citep[$z\sim0.6$;][]{Tremonti_etal:2007}.  This is suggestive of a
common quenching mechanism for massive PSBs that is simply more frequent at $z > 1$.  This is consistent with
a scenario in which {\em rapid quenching}, which is required to trigger the PSB phase, becomes more prevalent
at $z > 1$ \citep[see e.g.][]{Barro_etal:2013, Carnall_etal:2018, Belli_etal:2018}.

To establish the potential role of these high-velocity outflows in quenching star formation, it is useful to
consider the escape velocity $v_{\rm e}$ of the host galaxy
\begin{equation}
v_{\rm e} = \sqrt{\frac{2GM_{\rm d}}{r}}.
\end{equation}
For our high-$z$ PSBs, we use their dynamical mass $M_{\rm d}$ estimates (see
Section~\ref{velocity dispersions}), and find that $v_{\rm e}$ is typically $\sim950\rm\,km\,s^{-1}$
(determined at a galactocentric radius $r = 1\rm\,kpc$).  Since the high-velocity outflows detected in our
high-$z$ PSBs ($v_{\rm out}\sim1150\rm\,km\,s^{-1}$) will likely correspond to scales (i.e.\ radii) greater
than $1\rm\,kpc$, we find that $v_{\rm out} > v_{\rm e}$ for these galaxies.  This suggests the outflowing
gas will ultimately escape from the galaxy's gravitational well, or sweep into the surrounding
circum-galactic medium (CGM) creating an expanding shell (or bubble) that prevents future gas accretion
\citep[see e.g.][]{Lochhaas_etal:2018}.  For our high-$z$ PSBs, estimates of the time elapsed since starburst
$\Delta t_{\rm\,burst}$ are typically up to $\sim1\rm\,Gyr$ (Wild et al.\ in preparation).  Consequently, if
the outflowing winds were launched during the starburst event, they should have reached scales of several
hundred kpc by the time of our observations.  On these scales, one might expect the outflowing wind to have
encountered significant gas in the CGM and slowed down.  We make two main comments on this issue below.
\begin{enumerate}
\item For our high-$z$ PSBs, the stacked Mg\,{\sc ii} profile has an outflowing component with an absorption
depth $A_{\rm d}$ that could indicate the high-velocity outflows are only present in $\sim50$ per cent of our
sample (see Fig.~\ref{blue stack} and Section~\ref{Outflows}).  If this outflow signal is driven by the
youngest PSBs (as tentively indicated by our $D_{4000}$ analysis above), then the detected outflows might
not have necessarily reached the scales implied by the typical $\Delta t_{\rm\,burst}$ of the full sample.
\vspace{0.10cm}
\item Recent observations have indicated that both local and intermediate-redshift ($z\sim0.7$) PSBs retain a
significant molecular gas reservoir following the quenching of star formation \citep{French_etal:2015,
Rowlands_etal:2015, Suess_etal:2017}.
Consequently, if an AGN was triggered during the starburst event, it could linger into the post-starburst
phase and continue to drive the high-velocity outflows from the residual ISM, even after the quenching of
star formation.  It is also possible that the high-velocity wind is maintained during the post-starburst
phase by flickering AGN activity, or residual AGN activity that is optically obscured (so not detectable from
line emission in our spectra).
\end{enumerate}

With respect to the quenching mechanism, using the $K$-band structural parameters available within the UDS
field \citep{Almaini_etal:2017}, we find that the high-$z$ PSBs in our spectroscopic sample are typically
compact and spheroidally-dominated (effective radius $r_{\rm e}\sim2.1\rm\,kpc$ and S{\'e}rsic
index~$n\sim3.25$; see Table~\ref{sample properties}), with structures similar to that of our high-$z$
passive galaxies.  This result is consistent with previous studies that use only photometrically selected PSB
samples \citep[e.g.][]{Whitaker_etal:2012a, Yano_etal:2016, Almaini_etal:2017, Maltby_etal:2018}.  We note
that the $K$-band structural parameters in the UDS are from ground-based imaging, but consistent results are
also obtained using the limited fraction ($\sim20$ per cent) of our spectroscopic sample that has structural
parameters available from the {\em Hubble Space Telescope} ({\em HST}) $H$-band imaging of the CANDELS survey
\citep{Grogin_etal:2011, Koekemoer_etal:2011, vanderWel_etal:2012}.  Using the $K$-band structural
parameters, we also find a tentative hint that high-velocity outflows are more significant (as determined by
an F-test; see Section~\ref{Outflows}) in the stacked spectra of the most compact PSB galaxies
($r_{\rm e} < 2\rm\,kpc$).  Larger samples are needed to confirm these findings.  Taken together, these
results suggest a scenario involving a recent compaction event for high-$z$ PSBs, which may have triggered
the preceding starburst, high-velocity outflows and subsequent quenching.  Such an event could be, for
example, a gas-rich major merger \citep[e.g.][]{Hopkins_etal:2009, Wellons_etal:2015} or a dissipative
`protogalactic disc collapse' \citep[e.g.][]{Dekel_etal:2009, Zolotov_etal:2015}.

To build on this discussion, we also consider a sub-sample of spectroscopic PSBs (Spec+PCA) where photometric
PCA class (i.e.~{\em supercolour}) has been used to identify systems that are likely to have experienced a
more significant starburst (see Section~\ref{Sample selection}).  With respect to their structural
parameters, interestingly, we find that these Spec+PCA PSBs are significantly more compact
($r_{\rm e}\sim1.2\rm\,kpc$) than those defined from spectroscopy alone, and consistent with previous studies
that use only photometric PCA (supercolour) classifications \citep{Almaini_etal:2017, Maltby_etal:2018}.
This suggests that the addition of photometric-selection criteria isolates PSBs that have undergone a more
significant gas-rich dissipative event, which would also result in a more significant starburst, prior to
quenching.  Interestingly, we also find a tentative hint that high-velocity outflows are more significant in
these Spec+PCA PSBs.  If confirmed, this would suggest an intrinsic link between the compaction event,
subsequent starburst and the launch of high-velocity outflows.

Overall, our results suggest that for massive high-$z$ PSBs, high-velocity winds were launched during the
preceding starburst, and potentially represent the residual signature of a feedback process that quenched
their star formation.  These winds could either be caused by the starburst itself, or an AGN that was
triggered during the compaction event.  Within the optical regime probed by our spectra, several AGN
signatures are covered (e.g.\ [Ne\,{\sc v}] $\lambda\lambda\,3427$,\,$3581$\,\AA; [Ne\,{\sc iii}]
$\lambda3869$\,\AA).  However, we do not find any evidence for such features in our stacked PSB spectra (see
Figs.~\ref{red stack} and \ref{blue stack}), or on an individual basis.  Although, we note that due to the
[O\,{\sc ii}] condition used in our PSB criteria ($W_{\rm\,[O\,II]} > -5$\,\AA; see
Section~\ref{Sample selection}), we would likely remove any PSBs with AGN that cause significant optical
line emission \citep{Yan_etal:2006}.  Taken together, this suggests that for our high-$z$ PSBs either
i)~these galaxies were quenched via stellar feedback from the starburst itself; or ii)~if AGN feedback is
responsible, the AGN episode that triggered quenching does not linger into the post-starburst phase, as
required by some models \citep[e.g.][]{Hopkins:2012}.  We note, however, that using X-ray data the presence
of hidden AGN has been detected in galaxies where no optical AGN signatures are apparent
\cite[e.g.][]{Cimatti_etal:2013}.  Consequently, this issue will be explored in more detail in a forthcoming
paper, using the X-ray data available in the UDS field (Almaini et al., in preparation).

\section[]{Conclusions}

\label{Conclusions}

In this study, we have examined the prevalence of galaxy-scale outflows in post-starburst galaxies at high
redshift ($1 < z < 1.4$), using the deep optical spectra available in the UDS field.  Using a stacking
analysis, we find that for massive ($M_* > 10^{10}\rm\,M_{\odot}$) PSBs at $z > 1$, there is clear evidence
for a strong blue-shifted component to the Mg\,{\sc ii} absorption feature, indicative of high-velocity
outflows ($v_{\rm out}\sim1150\rm\,km\,s^{-1}$) in their interstellar medium.  These outflowing winds are
likely to have been launched during the preceding starburst, and therefore may represent the residual
signature of a feedback event which quenched their star-formation.  Using full spectral fitting, we also
obtain a typical stellar velocity dispersion $\sigma_*$ for these PSBs of $\sim200\rm\,km \,s^{-1}$, which
confirms they are intrinsically massive in nature (dynamical mass $M_{\rm d}\sim10^{11}\rm\,M_{\odot}$).
Given that these high-$z$ PSBs are also exceptionally compact ($r_{\rm e}\sim1$--$2\rm\,kpc$) and spheroidal
(S{\'e}rsic index~$n\sim3$), we propose that the outflowing winds may have been launched during a recent
compaction event (e.g.\ major merger or disc collapse) that triggered either a centralised starburst or AGN
activity.  Furthermore, we find no optical signatures of AGN activity in these galaxies, suggesting they were
either rapidly quenched by stellar feedback from the starburst itself, or that if AGN feedback is
responsible, the AGN episode that triggered quenching does not linger into the post-starburst phase.

\section[]{Acknowledgements}

We thank the anonymous referee for their detailed and insightful comments on the original version of this
manuscript, which helped to improve it considerably.  We also thank Mike Merrifield and Paula Coelho for
useful discussions.  This work is based on observations from ESO telescopes at the Paranal Observatory
(programmes 094.A-0410, 180.A-0776 and 194.A-2003).  AC acknowledges the support from grants PRIN-MIUR 2015,
ASI n.I/023/12/0 and ASI n.~2018-23-HH.0.


\bibliographystyle{mnras} \bibliography{DTM_bibtex}

\appendix

\section{Outflow velocity measurements}

\label{appendix A}

In this Appendix, we provide a simple calibration of the velocity offsets $\Delta{v}$ returned by our
two-component Mg\,{\sc ii} absorption model (i.e.\ systemic absorption + outflow; see
Section~\ref{Outflows}), so we can estimate the typical (i.e.\ median) outflow velocity
$\widetilde{v}_{\rm out}$ for our galaxy populations.  To achieve this, we run our fitting procedure on a
large number of simulated Mg\,{\sc ii} profile stacks, generated with a wide range of typical outflow
velocities $\widetilde{v}_{\rm out}$.  To generate our simulated Mg\,{\sc ii} profiles, we use the following
procedure.
\begin{enumerate}
\item We generate $2000$ Mg\,{\sc ii} profiles, each of which is modelled as a Gaussian-convolved doublet,
with a fixed intensity ratio of 1.2:1, as observed for massive high-$z$ galaxies \citep{Weiner_etal:2009}.
Each line has an intrinsic width matched to that of our observations ($\sigma_{\rm obs} = 8\,$\AA).  Random
noise is then added to these profiles at the required level
[$\rm S/N(\lambda_{\rm rest}\sim2800$\,\AA$)\sim5$; see Section~\ref{Composite spectra}].
\vspace{0.1cm}
\item We make an assumption on the contribution from both the systemic absorption and the outflowing
component to the overall stacked Mg\,{\sc ii} profile.  In this study, all our stacked spectra have an
Mg\,{\sc ii} absorption profile that exhibits a dominant systemic component (see~Fig.~\ref{blue stack}),
This is the case, even if an outflowing component is detected (e.g.\ high-$z$ PSBs).  There are several
potential explanations for this observation, e.g.\ outflows that are not in the line-of-sight and/or stellar
Mg\,{\sc ii} absorption (see Section~\ref{Outflows}, for further details).  Consequently, we require a large
fraction of our simulated Mg\,{\sc ii} profiles to be centred at the systemic absorption
(i.e.\ non-outflowing) in order to match our observations.  We assume that in any stacked Mg\,{\sc ii}
profile, $\sim50$ per cent of the input spectra will have an outflowing Mg\,{\sc ii}.  For these cases, the
relevant $\Delta\lambda$ is determined from an outflow velocity $v_{\rm out}$, which is randomly sampled from
a uniform distribution (range $0$--$v_{\rm max}$).
\vspace{0.1cm}
\item We then create a median stack of all $2000$ Mg\,{\sc ii} profiles (i.e.\ both those with systemic and
outflowing Mg\,{\sc ii}).  This simulated stacked Mg\,{\sc ii} profile will have an outflowing component with
a typical outflow velocity $\widetilde{v}_{\rm out}$, which is the median($v_{\rm out}$) of the input
profiles with outflowing Mg\,{\sc ii}.
\vspace{0.1cm}
\item We repeat the above procedure for various values of $v_{\rm max}$ \mbox{($0$--$4000\rm\,km\,s^{-1}$)},
in order to generate simulated Mg\,{\sc ii} profiles with a wide range of typical outflow velocities
($\widetilde{v}_{\rm out}$).
\end{enumerate}

\vspace{0.25cm}

\begin{figure}
\includegraphics[width=0.49\textwidth]{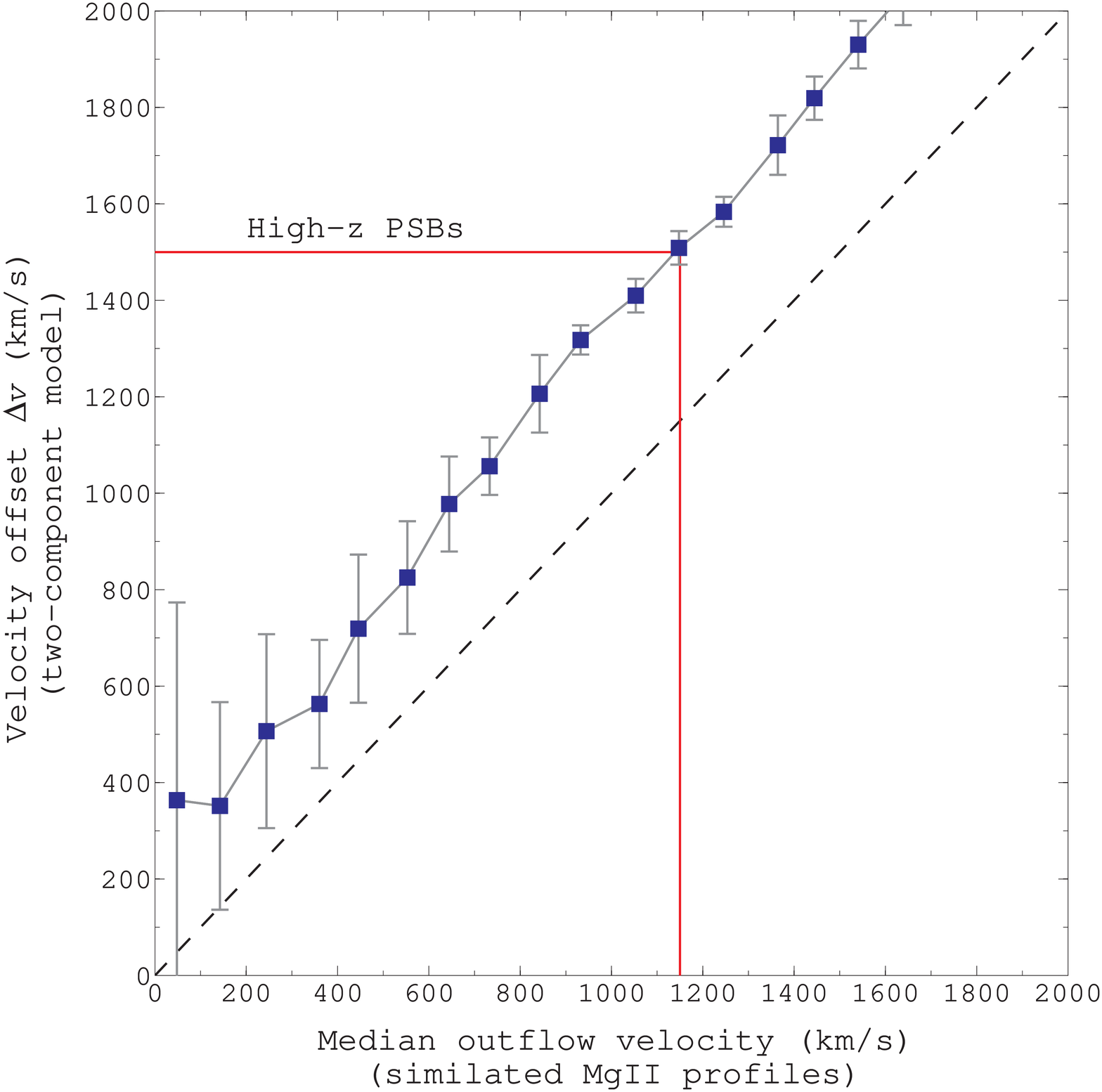}\\
\centering
\vspace{-0.1cm}
\caption{\label{simulations} The calibration of our measured outflow velocities.  A comparison of the typical
(i.e.\ median) outflow velocity of our simulated Mg\,{\sc ii} profiles $\widetilde{v}_{\rm out}$ with the
velocity offset $\Delta{v}$ returned by our two-component model.  The data have been binned by the median
outflow velocity $\widetilde{v}_{\rm out}$, and the error-bars represent the $1\sigma$ scatter in each bin.
The dashed line shows the 1:1 relation.  We find that for our two-component model, the outflowing component
has a velocity offset $\Delta{v}$ which systematically over-estimates the median outflow velocity
$\widetilde{v}_{\rm out}$ by $\sim350\rm\,km\,s^{-1}$.   For our high-$z$ PSBs, our two-component model
yields an outflowing component with a velocity offset $\Delta{v}\sim1500\pm150\rm\,km\,s^{-1}$ (red line).
Therefore, these simulations suggest that the true outflow velocity is actually
$\widetilde{v}_{\rm out}\sim1150\pm160\rm\,km\,s^{-1}$.}
\end{figure}

For each simulated Mg\,{\sc ii} profile, we use our two-component model (see Section~\ref{Outflows}) in order
to estimate the velocity offset~$\Delta{v}$ of the outflowing component.  In Fig.~\ref{simulations}, we
present a comparison of the median velocity of our simulated Mg\,{\sc ii} profiles $\widetilde{v}_{\rm out}$
with the velocity offset $\Delta{v}$ returned by our two-component model.  This reveals that our
two-component model yields outflowing components with a velocity offset $\Delta{v}$ that systematically
over-estimates the typical (i.e.\ median) outflow velocity $\widetilde{v}_{\rm out}$ by
$\sim350\rm\,km\,s^{-1}$.  For our high-$z$ PSBs, our two-component model yields an outflowing component with
a velocity offset $\Delta{v}\sim1500\pm150\rm\,km\,s^{-1}$ (see Section~\ref{Outflows}).  Therefore, these
simulations indicate that, in this case, the true outflow velocity is actually typically
$\widetilde{v}_{\rm out}\sim1150\pm160\rm\,km\,s^{-1}$.  The uncertainty in this $\widetilde{v}_{\rm out}$
estimate has been determined by combining the errors in $\Delta{v}$ and those of these simulations, in
quadrature.

\section{Stellar Mg\,{\sc ii} absorption}

\label{appendix B}

For PSB and passive galaxies, the dominant contribution to their systemic Mg\,{\sc ii} absorption is expected
to be stellar in origin.  In this work, we have modelled the systemic absorption (ISM + stellar) as a single
component comprising a Gaussian-convolved doublet (see Section~\ref{Outflows}).  An alternative approach is
to use synthetic stellar libraries, to estimate and remove the stellar Mg\,{\sc ii} component using full
spectral fitting.  In this Appendix, we explore this alternative approach and its impact on our results.

To obtain spectral fits covering the Mg\,{\sc ii} region ($\lambda\sim2800$\,\AA), it is necessary to use
synthetic (i.e.\ theoretical) stellar libraries.  We note that these libraries are not as robust as the
empirical libraries available in the optical regime \citep[e.g.\ MILES;][]{Vazdekis_etal:2010}, but are
nonetheless useful to estimate the UV stellar continuum.  One well-established library is UVBLUE
\citep{Rodriguez_etal:2005}, a high-resolution ($R=50\,000$) theoretical stellar library of $1770$ stars,
which covers all spectral types and spans a wide range in temperature, metallicity and surface gravity.  For
our stacked spectra (blue-optimised; Fig.~\ref{blue stack}), we use the UVBLUE library to perform full
spectral fitting via {\sc ppxf} \citep{Cappellari&Emsellem:2004, Cappellari:2017}.  These fits are performed
over a wide wavelength range ($2550$--$4350$\,\AA)\ and assume solar metallicity, as expected for the stellar
metallicity of massive galaxies at $z>1$ \citep[e.g.][]{Sommariva_etal:2012}.  In the fitting, we also
include an additive polynomial correction to amend the continuum shape for e.g.\ dust effects, mismatches
between the model and data, and spectrophotometric inaccruacies.  This is necessary in order to obtain an
adequate fit to the continuum on either side of the Mg\,{\sc ii} feature.  For the fits presented below, this
correction is a Legendre polynomial $P_{n}(x)$ of degree $n = 20$.  We note, however, that consistent
conclusions are also obtained using a correction with degree $n = 1$ (i.e.\ a linear correction).

\begin{figure*}
\includegraphics[width=0.95\textwidth]{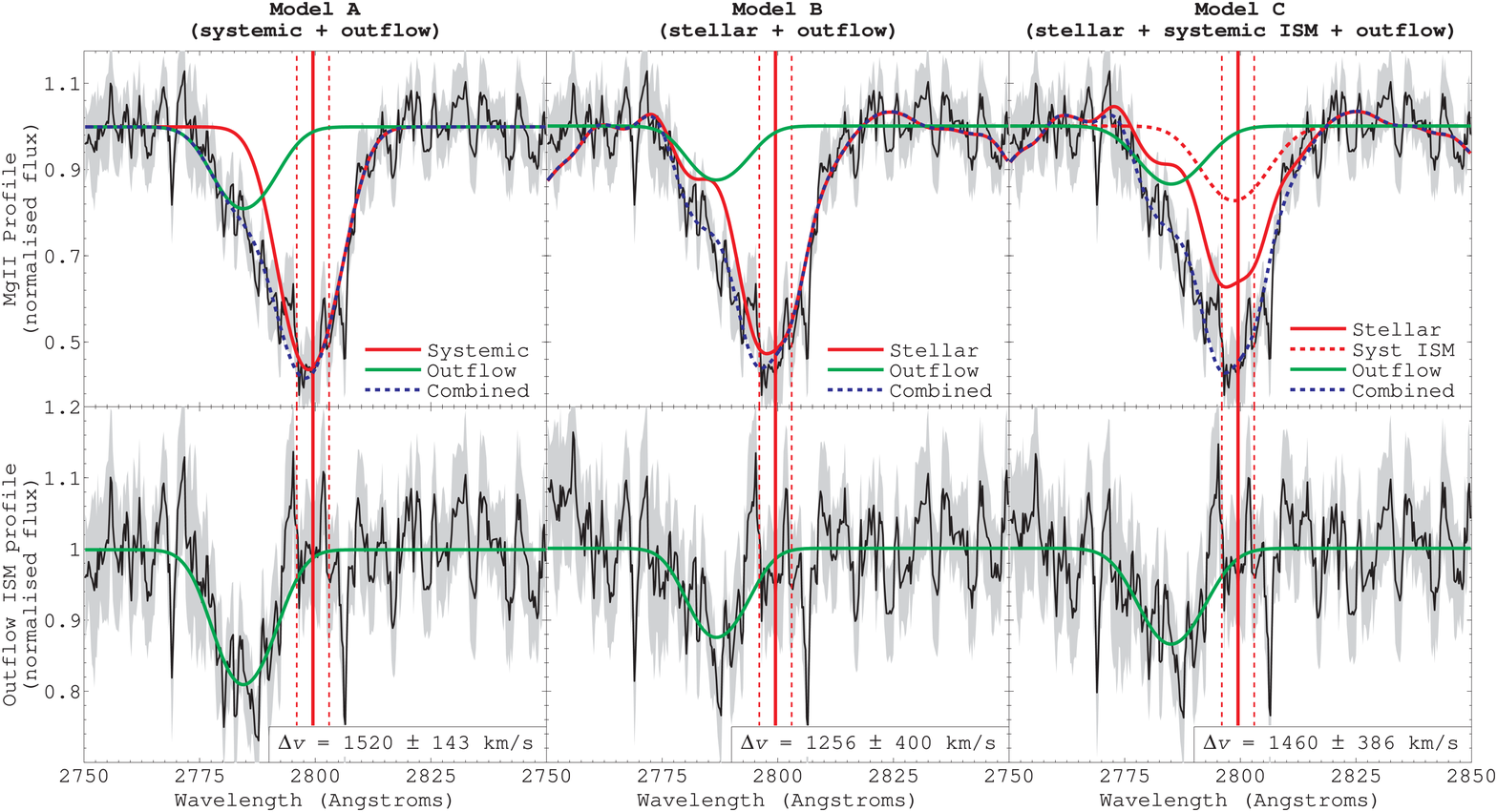}\\
\centering
\vspace{-0.0cm}
\caption{\label{stellar component}A comparison of various models for the stacked Mg\,{\sc ii} absorption
profile of our high-$z$ PSBs (see Fig.~\ref{blue stack}).  Left-hand panels: the two-component model
(systemic absorption + outflow) presented in Section~\ref{Outflows}, which consists of two Gaussian-convolved
doublets (Model~A).  Centre panels: an alternative two-component model (stellar + outflow), where the
blue-side of the Mg\,{\sc ii} profile was masked in the fit to the stellar continuum (Model~B).  Right-hand
panels: a three component model (stellar + systemic ISM + outflow), where the full Mg\,{\sc ii} profile was
masked in the fit to the stellar continuum (Model~C).  In each case, we show the decomposition of the stacked
Mg\,{\sc ii} profile (top panels), and the residual outflowing ISM profile, i.e.\ after the removal of the
systemic component (bottom panels).  We note that in the fits to the stellar continuum, a weak unknown
absorption feature is revealed blue-ward of Mg\,{\sc ii}, potentially accounting for some of the profile's
asymmetric nature.  Nonetheless, the presence of an outflowing component is still required in all models
(significance $>3\sigma$).  Furthermore, consistent velocity offsets are also obtained with each model
($\Delta{v}\sim1500\rm\,km\,s^{-1}$).  The $1\sigma$ uncertainties in the velocity measurements are
determined using the variance between analogous fits performed on $1000$ simulated spectra generated via a
bootstrap analysis.}
\end{figure*}

For our high-$z$ PSBs, we use a full spectral fit to normalise the stacked spectrum with respect to the
stellar component, and hence determine the Mg\,{\sc ii} profile of the ISM absorption (see
Fig.~\ref{stellar component}).  Depending on the nature of the spectral fit used, we then model the ISM
component with one or two components, as required (see details below).  In the following, we compare two
resultant models for the Mg\,{\sc ii} profile with that of our original two-component model
(see~Section~\ref{Outflows}).  We define the three different models as follows:
\begin{enumerate}
\item {\em Model A}: the two-component model (systemic absorption + outflow) presented in
Section~\ref{Outflows} (see Fig.~\ref{blue stack}).  This consists of two Gaussian-convolved doublets: one
fixed at the rest-frame wavelength for the systemic absorption (stellar + ISM), and another with a free
centroid to model the outflow.
\vspace{0.1cm}
\item {\em Model B}: a two-component model (stellar + outflowing ISM), with a full spectral fit used to
determine the stellar component.  To avoid the influence of the outflowing ISM absorption on the fit, the
blue-side of the Mg\,{\sc ii} profile is masked ($\lambda < 2800$\,\AA).  The red-side of the Mg\,{\sc ii}
profile ($\lambda > 2800$\,\AA) is unmasked, which essentially forces the fit to account for all the systemic
absorption with the stellar component.  In this case, only a single outflowing component
(i.e.\ Gaussian-convolved doublet) is required to account for the ISM absorption.
\vspace{0.1cm}
\item {\em Model C}: a three-component model (stellar + systemic ISM + outflowing ISM).  In this case, a full
spectral fit is used to determine the stellar component, but in the fitting the entire Mg\,{\sc ii} profile
is masked ($2760 < \lambda < 2830$\,\AA).  Since our PSBs are expected to retain a significant residual ISM
component (see Section~\ref{outflow origins}), we do expect some ISM contribution to the systemic absorption.
Therefore, masking the entire Mg\,{\sc ii} profile avoids forcing the fit to account for all the systemic
absorption with the stellar component.  In this case, we require two components to account for the ISM
absorption (systemic + outflowing), each of which consists of a Gaussian-convolved doublet. 
\end{enumerate}

In Fig.~\ref{stellar component}, we present and compare the results of each model for the stacked
Mg\,{\sc ii} profile of our high-$z$ PSB galaxies.  In the full spectral fits to the stellar component
(Models B and C), we note that the stellar continuum has revealed a potential weak feature blue-ward of
Mg\,{\sc ii}, that could potentially account for some of the asymmetric nature.  We have thoroughly explored
this issue in the UVBLUE stellar models and found that this feature is only important in the atmospheres of
F-stars, potentially due to a weak molecular absorption line that becomes significant at those effective
temperatures ($5000 < T_{\rm eff} < 7500$\,K).  Similar results are also found using an alternative synthetic
stellar library provided by \cite{Coelho_etal:2014}.  Nonetheless, despite this issue, we find that for both
models explored (Models B and C) the Mg\,{\sc ii} profile still presents a significant excess of blue-shifted
absorption with respect to the stellar component.  Furthermore, in comparison to our original two-component
model (Model A), we find that i)~although weakened, the significance of an outflowing ISM component remains
$>3\sigma$ (as determined by an F-test); and ii)~consistent velocity offsets for the outflow are obtained
($\Delta{v}\sim1500\rm\,km\,s^{-1}$).

With respect to our stellar continuum fits, we note that the synthetic stellar libraries used are based on
theoretical models that are known to suffer from various issues.  For example, it is known that i)~various
regions of the UV spectrum are poorly reproduced; and ii)~the prominent metallic lines in F/G stars
(including Mg\,{\sc ii}), are always stronger in the synthetic spectra than in observed stars \citep[see][for
further details]{Rodriguez_etal:2005}.  This is also true to a lesser extent for A-stars \citep[see
fig.~10 from ][]{Rodriguez_etal:2005}.  We also note that in our models, the strength of the stellar
Mg\,{\sc ii} absorption differs depending on the nature of the Mg\,{\sc ii} mask used in the fit. This
implies that the Mg\,{\sc ii} feature is essentially unconstrained in our fits from other features in the
blue-optical spectrum ($2550$--$4350$\,\AA).  Due to these issues, it is likely that the unknown weak
feature on the blue-side of Mg\,{\sc ii} is also unconstrained and therefore uncertain in nature.  These
issues require further exploration in order to ascertain the true importance of this feature in the stellar
models.  Nonetheless, even in the case where the entire systemic absorption is accounted for by the stellar
component (Model B), an outflowing ISM component is still required to account for the stacked Mg\,{\sc ii}
profile. Therefore, we conclude that our detection of high-velocity outflows in high-$z$ PSBs is robust to
the nature of the stellar Mg\,{\sc ii} component.

\bsp

\label{lastpage}

\end{document}